\documentclass[twocolumn,showpacs,fleqn,nobibnotes]{revtex4}

\usepackage{amsmath}
\usepackage{graphicx}
\usepackage{float}
\usepackage{subfigure}

\def\lsim{\raise0.3ex\hbox{$<$\kern-0.75em\raise-1.1ex\hbox{$\sim$}}}
\def\gsim{\raise0.3ex\hbox{$>$\kern-0.75em\raise-1.1ex\hbox{$\sim$}}}

\begin{document}
\newcommand\ie {{\it i.e.}}
\newcommand\eg {{\it e.g.}}
\newcommand\etc{{\it etc.}}
\newcommand\cf {{\it cf.}}
\newcommand\etal {{\it et al.}}
\newcommand{\jp}{$ J/ \psi $}
\newcommand{\pp}{$ \psi^{ \prime} $}
\newcommand{\ppp}{$ \psi^{ \prime \prime } $}
\newcommand{\dd}[2]{$ #1 \overline #2 $}
\newcommand\noi {\noindent}
\def\beq{\begin{equation}}
\def\eeq{\end{equation}}
\def\beqa{\begin{eqnarray}}
\def\eeqa{\end{eqnarray}}
\newcommand{\coes}{\c{c}\~oes }
\newcommand{\cao}{\c{c}\~ao }
\newcommand{\sao}{s\~ao }
\newcommand{\ii }{\'{\i}}
\newcommand{\nnn}{\noindent}
\newcommand{\shat}{\hat{s}}
\newcommand{\that}{\hat{t}}
\newcommand{\uhat}{\hat{u}}
\newcommand{\thetahat}{\hat{\theta}}
\newcommand{\rr}{\mbox{\boldmath $r$}}
\newcommand{\rrn}{\mbox{$r$}}
\newcommand{\rp}{\mbox{\boldmath $p$}}
\newcommand{\rqq}{\mbox{\boldmath $q$}}


\title{Gluon saturation and the Froissart bound: a simple approach}
\author{$^1$F. Carvalho, $^2$F.O. Dur\~aes, $^{3}$V.P. Gon\c{c}alves and $^1$F.S. Navarra}
\affiliation{$^1$Instituto de F\'{\i}sica, Universidade de S\~{a}o Paulo\\
C.P. 66318,  05315-97, S\~{a}o Paulo, SP, Brazil\\
$^2$Dep. de F\'{\i}sica, Centro de Ci\^encias e Humanidades,\\
Universidade Presbiteriana Mackenzie, C.P. 01302-907, S\~{a}o Paulo, SP,Brazil\\
$^3$Instituto de F\'{\i}sica e Matem\'atica, \\
Universidade Federal de Pelotas, Caixa Postal 354, CEP 96010-900, Pelotas, RS, Brazil \\
}


\begin{abstract}
At very high energies we expect that the hadronic cross sections
satisfy the Froissart bound, which is a well-established property of
the strong interactions. In  this energy regime we also expect the
formation of the Color Glass Condensate, characterized by  gluon
saturation and a typical momentum scale: the saturation scale $Q_s$.
In this paper we show that if a saturation window exists between
the nonperturbative and perturbative regimes of  Quantum
Chromodynamics (QCD), the total cross sections satisfy the Froissart
bound. Furthermore, we show that our approach allows us to
describe the high energy experimental data on  $pp/p\bar{p}$  
total cross sections.
\end{abstract}


\maketitle

\section{Introduction}

\begin{figure}[t]
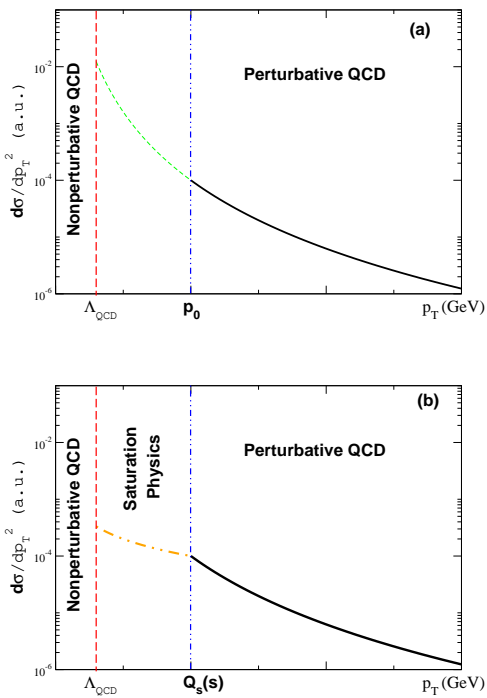

\includegraphics[scale=0.25]{dsdptns.eps}\\
\vspace{0.8cm}
\includegraphics[scale=0.25]{dsdptsat2.eps}
\caption{Schematic behavior of the momentum distribution in  the minijet model (a) 
and  in the model proposed in this paper (b). While in the minijet model the region 
$\Lambda_{QCD} \le p_T \le p_0$ is disregarded, the region 
$\Lambda_{QCD} \le p_T \le Q_s (s)$ is included in our model and its contribution 
to the total cross section increases when the energy rises.} 
\label{fig1}
\end{figure}

Understanding the behavior of high energy hadron reactions from a
fundamental perspective within Quantum Chromodynamics (QCD) is an
important goal of particle physics. Since  it was observed that the
total hadronic cross sections grow with the center of mass energy 
($\sqrt{s}$), much theoretical effort has been devoted to  explain  
this growth.
In particular, a QCD based explanation for this rising behavior was
proposed by Gaisser and Halzen \cite{halzen}: the cross section
would grow because partons would start to play a role in the hadronic
reactions. At higher energies, lower values of the Bjorken $x$ are
accessible and the parton distributions (especially the gluon
distribution) grow very rapidly leading to rising cross sections.
The basic idea of this approach, called minijet model, is that the
total cross section can be decomposed as follows (see Fig.
\ref{fig1} (a)):
\begin{eqnarray}
\sigma_{tot}   = \int_0^{p^2_{0}} d p^2_T \frac{d \sigma}{d p^2_T}
+ \int_{p_{0}^2}^{s/4} d p^2_T \frac{d \sigma}{d p^2_T} = \sigma_0
 +   \sigma_{pQCD} \label{sigsum1}
\end{eqnarray}
where $\sigma_0$ characterizes the nonperturbative contribution,
which is in general taken as energy-independent at high energies \cite{dosch},
and $\sigma_{pQCD}$ is calculable in perturbative QCD with the use of an
arbitrary cutoff at low transverse momenta $p_0$. Unfortunately,
this approach implies  a power-like energy behavior for the total
cross section, violating the Froissart bound, which is a
consequence of the unitarity of the $S$ matrix and states that total
cross sections cannot grow faster than $\ln^2 s$ as $s \rightarrow
\infty$. This bound is a well-established property of the strong
interactions and puts a strict limit on  the rate of growth with
energy of the total cross sections. 
Over the years several solutions have been proposed to cure the too  fast
growth found  in the minijet model \cite{halzen}, most of them using the
eikonal formalism \cite{durand}. 

In parallel with these developments
the study of the high energy limit of the linear evolution equations
(DGLAP and BFKL) \cite{barone} revealed that they  should be
modified and gluon recombination effects (consequence of the high
density of gluons) should be included in the QCD evolution \cite{glr}.
This expectation can be easily understood: while for large momentum
transfer $k_{\perp}$, the BFKL equation predicts that the mechanism
$g \rightarrow gg$ populates the transverse space with a large
number of small size gluons per unit of rapidity (the transverse
size of a gluon with momentum $k_{\perp}$ is proportional to
$1/k_{\perp}$), for small $k_{\perp}$ the produced gluons overlap
and fusion processes, $gg \rightarrow g$, are equally important.
Currently, one believes that the small-$x$ gluons in a hadron
wave function should form a Color Glass Condensate (CGC) which is
described by an infinite hierarchy of coupled evolution equations
for the correlators of Wilson lines \cite{cgc,bk}. This new state of
matter is characterized by gluon saturation and by a typical momentum
scale, the saturation scale $Q_s$, which grows with the energy  and
determines the critical line separating  the linear
 and saturation regimes of the QCD
dynamics. The saturation effects are small for $k_{\perp}^2 >
Q_{\mathrm{s}}^2$ and very strong for $k_{\perp}^2 <
Q_{\mathrm{s}}^2$. Experimentally, there are strong evidences of  
nonlinear (saturation) effects at DESY-HERA. In particular, the DESY
$ep$ HERA data in the small-$x$ and low-$Q^2$ region can be
successfully described in terms of saturation models
\cite{satmodels,kkt,kgn1,iim,IANCUGEO,GBW,dhj}, with the measured cross sections presenting
the geometric scaling property \cite{scaling}, which is an intrinsic
property of the CGC physics. Moreover, the CGC physics is able to
describe quite well the $d Au$ RHIC data (see, e.g. Ref.
\cite{rhic}). These results give strong support to the existence of
a saturation regime in the QCD dynamics (for recent reviews see,
e.g., Ref. \cite{cgc}).

Some attempts  to  reconcile the QCD parton picture with 
the Froissart limit using saturation physics were proposed in recent years, 
 but the question remains open \cite{iancu,kovner,mclerran}. In this
paper we propose a very simple phenomenological approach to treat
this problem.  In the next section we briefly describe the minijet model and 
how we include saturation effects in it. In the subsequent sections we present 
our numerical results and discuss them.

\section{The minijet model with saturation}

In what follows 
we generalize the minijet model assuming
the existence of the saturation regime in the high energy limit. More precisely,  
we assume the existence of a saturation window between the
nonperturbative and perturbative regimes of QCD, which grows when
the energy  increases (since $Q_s$ grows with the energy). This
window is shown in  Fig. \ref{fig1} (b). We now  generalize Eq.  
(\ref{sigsum1}) introducing the saturation window:
\begin{eqnarray}
\sigma_{tot} &=&
\int_0^{\Lambda^2_{QCD}} d p^2_T \frac{d \sigma}{d p^2_T}  +
\int_{\Lambda^2_{QCD}}^{Q_s^2}  d p^2_T \frac{d \sigma}{d p^2_T} +
\int_{Q_s^2}^{s/4}  d p^2_T \frac{d \sigma}{d p^2_T}  \nonumber  \\
&=& \sigma_0 \, + \, \sigma_{sat} + \, \sigma_{pQCD}\,,
\label{sigsum2}
\end{eqnarray}
where the saturated component, $\sigma_{sat}$, contains the dynamics
of the interactions at scales lower than the saturation scale. In
this region nonlinear effects are strong, changing  the
$p_T$  behavior of the differential cross section, which becomes much 
less singular in the low $p_T$ region, as we can see in  Fig. \ref{fig1} (b). 
The simple QCD collinear factorization 
formulas do not in general apply in this
region.

\subsection{$\sigma_{pQCD}$}

At high transverse momenta we keep using the same collinear factorization formula
employed in \cite{halzen,durand} with the necessary update of the parton densities.  
The saturation scale arises as a natural cut-off at low transverse
momenta of the perturbative (minijet) cross section component,
$\sigma_{pQCD}$, which is now given by:
\beq
\sigma_{{pQCD}}= \frac{1}{2} \int_{Q_s^2} dp_T^2  \sum_{i,j} \int
dx_{1}\, dx_{2} \, f_i(x_1,p_T^2) \, f_j(x_2,p_T^2) \, 
\hat{\sigma}_{ij} 
\label{smjet} 
\eeq 
where $f_i(x,Q^2)$ is the
parton density of the species $i$ in the proton extracted from deep
inelastic scattering (DIS)  and $\hat{\sigma}_{ij}$ is the leading order 
elementary parton-parton cross section.  At very high energies the
cross section (\ref{smjet}) is dominated by gluon-gluon
interactions. In what follows we use the MRST leading order parton 
distributions \cite{mrst} in our calculations of $\sigma_{pQCD}$. 
Similar results are obtained
using for instance the CTEQ6-LO parton distribution sets \cite{cteq}.

In order to evaluate  $\sigma_{pQCD}$ we need to specify $Q_s$, which  is
determined by the solution of the nonlinear evolution equation
associated to CGC physics \cite{cgc,bk}. It is  given by:
\beq
Q_s^2 (x) = Q_0^2\,(\frac{x_0}{x})^{\lambda}  
\label{qsgbw}
\eeq
where $x$ is the
Bjorken variable, with $Q_0^2=0.3$ GeV$^2$ and $x_0=0.3 \times 10^{-4}$ fixed by the 
initial condition. The saturation exponent $\lambda$ has been estimated
considering different approximations for the QCD dynamics, being $
\approx 0.3$ at NLO accuracy \cite{trianta}, in  agreement with the
HERA phenomenology, where the parameters $Q_0$, $x_0$ and
$\lambda$ were fixed by fitting the  $ep$  HERA data \cite{satmodels}.

When  we go from deep inelastic scattering to hadron - hadron
collisions,  there  is some ambiguity in the definition of the
equivalent of the Bjorken $x$. Following \cite{bartels} we take $x$ to be  
\beq
x = \frac{q_0^2}{s} 
\label{xbj}
\eeq 
where $q_0$ is a momentum scale to be determined.  From (\ref{qsgbw}) and (\ref{xbj}) 
we immediately see that:
\beq
Q_s^2 (s) \propto s^{\lambda} 
\label{qsats}
\eeq
with the constant of proportionality being determined by  HERA data and our choice
of $q_0$. 

As it will be seen, in the high energy  limit  $\sigma_{pQCD}$ is the most important  
contribution to $\sigma_{tot}$. With a constant infra-red cut-off, as in 
(\ref{sigsum1}), it would grow too fast. 
The introduction of a cut-off increasing with 
energy can tame this growth, since the bulk of the integral in (\ref{smjet})  
comes from the low momentum region. 
This procedure was already employed, for example in 
\cite{kari}, in a purely pragmatic approach. Here we establish a connection between this 
cut-off and the energy behavior of $Q_s$, controlled by the parameter $\lambda$ in 
(\ref{qsats}). Our procedure is more physical and, at the same time, imposes restrictions 
on $\lambda$.

\subsection{$\sigma_{sat}$}

In order to calculate the total cross section we also need to
specify the saturated component. There are a few models for
$\sigma_{sat}$ \cite{bartels,shoshi,kppp}. Most of them are formulated in the color dipole
picture, in which the projectile proton is treated as a color
dipole, which interacts with the target proton. We shall use the
model proposed in Ref. \cite{bartels}, in which the total cross
section is given by: 
\beq \sigma_{sat}=
\int d^2 r
|\Psi_{p}(r)|^ 2 \sigma_{dip} (x,r) 
\label{sigsat}
\eeq 
where $r$ is the dipole transverse radius and   
the proton wave function $\Psi_p$ is chosen to be:
\begin{equation}
|\Psi _p(r)|^ 2 =
\frac{1}{2 \pi S^ 2_ p} \, exp\bigg(-\frac{r^ 2}{2 S^ 2 _p}\bigg)
\label{funcao_onda}
\end{equation}
with $S_p=0.74$ fm and the dipole-proton cross section reads:
\beq
\sigma_{dip}(x,r) = 2 \int d^2 b \, {\cal{N}}(x,r,b) = 
\overline{\sigma}  \,  {\cal{N}}(x,r)
\label{sigdip}
\eeq
where $\overline{\sigma} = 2 \pi R_p^2$, with $R_p = 0.9$ fm. 
The dipole scattering  amplitude,  ${\cal{N}}(x,r,b)$, should be given by the
impact parameter dependent solution of a non-linear evolution equation, such
as the Balitsky-Kovchegov equation \cite{bk}. A complete solution is not yet available and
we would have to use models for  ${\cal{N}}(x,r,b)$.  In \cite{bartels} it was 
assumed that ${\cal{N}}(x,r,b)$ falls exponentially with $b$. In this case 
$\sigma_{sat}$ does not violate the Froissart bound. However in \cite{kovner} it has been  
argued that the dipole amplitude decays only as a power of the impact parameter in the 
periphery of the proton and this dependence will, after integration in $b$,  lead to 
logarithmic divergences.  Here we prefer to avoid the use
of models and, instead, assume the factorization implied by the second equality in 
(\ref{sigdip}). With  this assumption we decouple the impact parameter and energy 
dependences and focus only on the energy behavior of the dipole and hadron-hadron cross 
sections. 

In the literature there are many parameterizations of the dipole amplitude. A brief discussion 
of the features of some recent ones can be found in \cite{rhic}. In what follows we shall use
two of them, which were shown to give a reasonable description of both HERA and RHIC data
\cite{rhic}. Most of the parameterizations follow the Glauber-like formula originally 
introduced by Golec-Biernat and W\"usthoff \cite{GBW}. The differences among them are in the
anomalous dimension, $\gamma$.  In the KKT model \cite{kkt}  the expression 
for the quark dipole-target forward scattering amplitude  is given by \cite{kkt}:
\beq\label{glauber2}
{\cal{N}}(r,x) \, = \, 1- \exp\left[-\frac{1}{4} \left(r^2
\, \bar{Q}_s^2\right)^{\gamma(Y,r^2)}\right].
\eeq
where  $\bar{Q}_s^2 = \frac{C_F}{N_c} \, Q_s^2$ and  the  anomalous dimension 
$\gamma(Y, r^2)$  is 
\beq\label{gamma}
\gamma(Y, r^2) \, = \, \frac{1}{2}\left(1+\frac{\xi 
(Y, r^2)}{\xi (Y,r^2) + \sqrt{2 \,\xi (Y, r^2)}+  7
\zeta(3)\, c} \right),
\eeq
with $c$ a free parameter ({ which was fixed in \cite{kkt} to $c=4$}) and 
\beq\label{xi}
\xi (Y, r^2) \, = \, \frac{\ln\left[1/( r^2 \, Q_{s0}^2 ) 
\right]}{(\lambda/2)(Y-Y_0)}\,.
\eeq
The authors of \cite{kkt} assume that the saturation scale can be expressed by $
Q_s^2(Y)  = \Lambda^2 A^{1/3} \left(\frac{1}{x}\right)^{\lambda}$. 
 The form of the anomalous
dimension is inspired by the analytical solutions to the BFKL equation. 
Namely, in the limit $r \rightarrow 0$ with $Y$ fixed we 
recover the anomalous dimension in the double logarithmic
approximation $\gamma \approx 1 - \sqrt{1/(2 \, \xi)}$. In another
limit of large $Y$ with $r$ fixed, Eq. (\ref{gamma}) reduces to the
expression of the anomalous dimension near the saddle point in the
leading logarithmic approximation $\gamma \approx
\frac{1}{2} + \frac{\xi}{14 \, c \, \zeta (3)}$. Therefore Eq. (\ref{gamma}) 
mimics the onset of the geometric scaling region \cite{iim,IANCUGEO}. 
In the calculations of  Ref. \cite{kkt} it is assumed that a 
 characteristic value of $r$ is $r \approx 1/(2 \, k_T)$ where $k_T$ is the 
transverse momentum of the valence 
quark and $\gamma$ was approximated by 
$\gamma(Y, r^2) \approx \gamma(Y,1/(4 \, k_T^2))$. In the above expressions the 
parameters $\Lambda=0.6$~GeV and 
$\lambda=0.3$ are fixed by DIS data \cite{GBW}. Moreover, the authors assume $Y_0 = 0.6$. 
The initial saturation scale used
in (\ref{xi}) is defined by $Q_{s0}^2=Q_s^2(Y_0)$ with $Y_0$ being  the
lowest value of rapidity at which the low-$x$ quantum evolution
effects are essential. As demonstrated in Ref. \cite{kkt} this parameterization is able to 
describe the $dAu$ RHIC data when the forward dipole cross section is convoluted with the 
respective fragmentation function and the parton distributions for the deuteron.

\begin{figure}[t]
\includegraphics[scale=0.65]{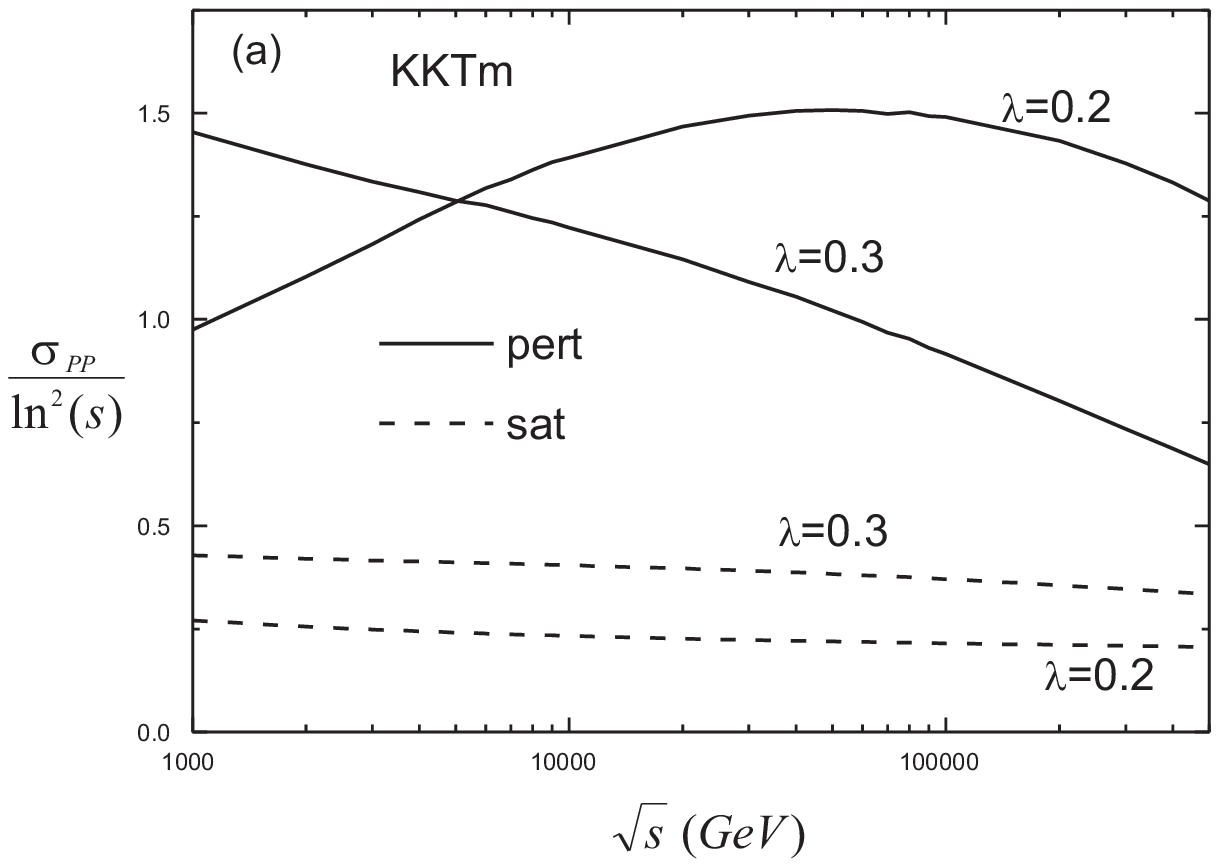}\\
\hspace{-0.6cm}
\includegraphics[scale=0.65]{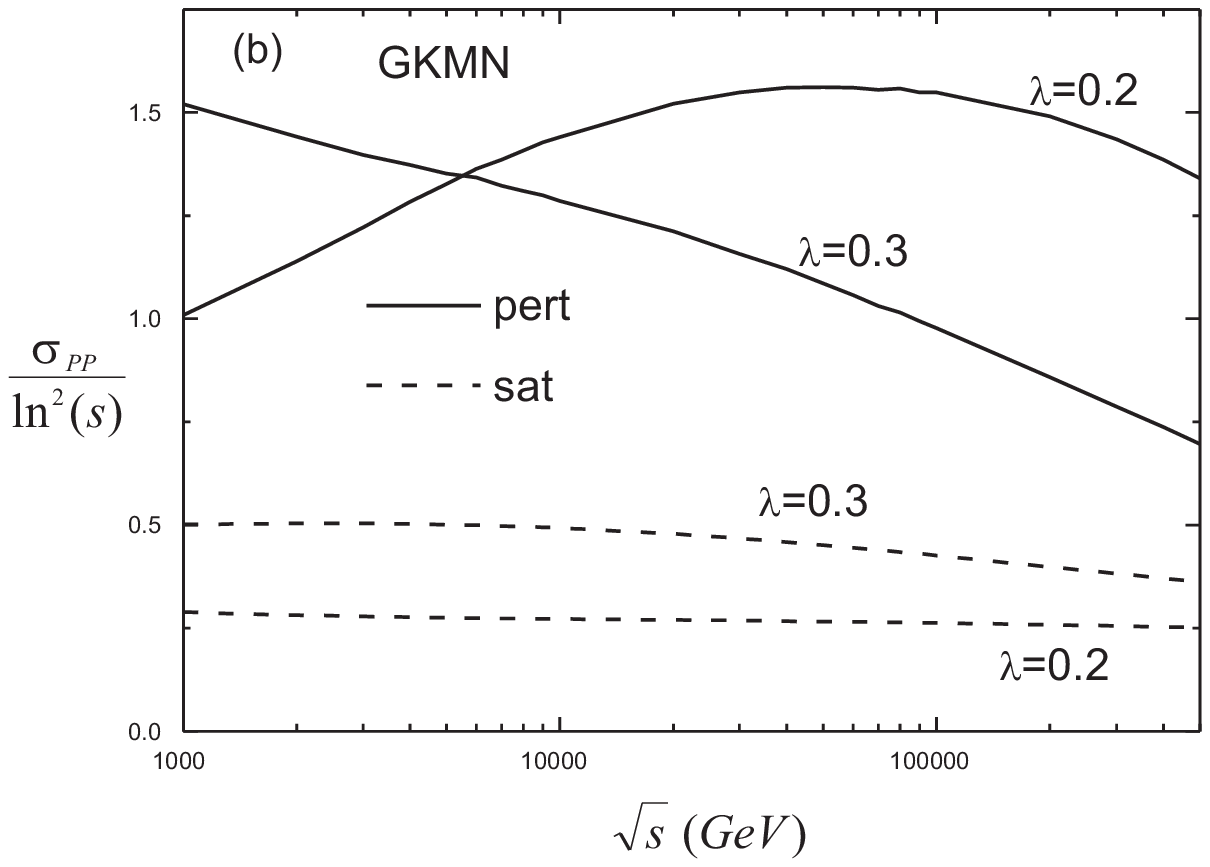}
\caption{Energy behavior of the perturbative (solid lines) and saturated components 
(dashed lines) of the total cross section (normalized by $\ln^2 s$ and in arbitrary units) 
for two different values of the exponent $\lambda$. In (a) and (b) we use the KKTm and 
GKMN dipole cross sections respectively. The perturbative component is the same.} 
\label{fig2}
\end{figure}

In Ref. \cite{dhj} another phenomenological saturation  model has been 
proposed in order to describe the $dAu$ RHIC data (hereafter denoted DHJ model). The basic 
modification with respect to the KKT model is the parameterization of the anomalous 
dimension 
which is now given by 
\beq\label{gamma_dhj}
\gamma(Y, r^2) \, = \gamma_s + \Delta \gamma (Y,r^2)
\eeq
where 
\beq
\Delta \gamma (Y,r^2) = (1 - \gamma_s) \frac{ |\log\frac{1}{r^2 Q_T^2}|}{\lambda Y +
 |\log\frac{1}{r^2 Q_T^2}| + d\sqrt{Y}}\,\,\,,
\eeq
with $Q_T = Q_s(Y)$ a typical hard scale in the process, $\lambda = 0.3$ and $d = 1.2$. 
Moreover, $\gamma_s = 0.63$ is the anomalous dimension of the  BFKL evolution with 
saturation  boundary condition.  Similarly to the KKT model this model is 
able to describe the $dAu$ RHIC data.

As already discussed in Ref. \cite{kgn1}, based on 
the universality of the hadronic 
wave function predicted by the CGC formalism,  we might expect that the KKT and DHJ 
parameterizations would also describe the HERA 
data on  proton structure functions  in the  kinematical 
region where the saturation effects should be present (small $x$ and low $Q^2$). 
However, as shown in \cite{rhic}, this is not the case and neither KKT nor DHJ give 
an acceptable description of the HERA data on $F_2$.

\begin{figure}[t]
\includegraphics[scale=0.65]{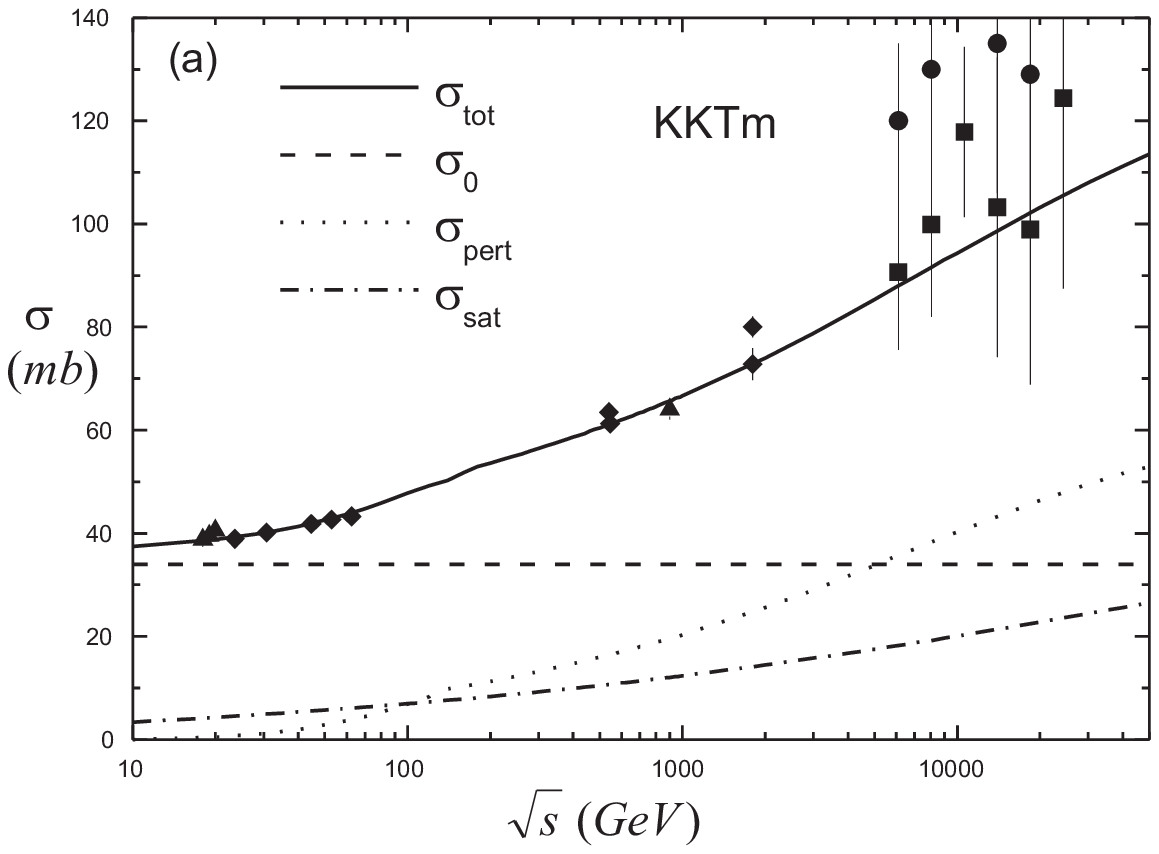}\\
\hspace{-0.1cm}
\includegraphics[scale=0.65]{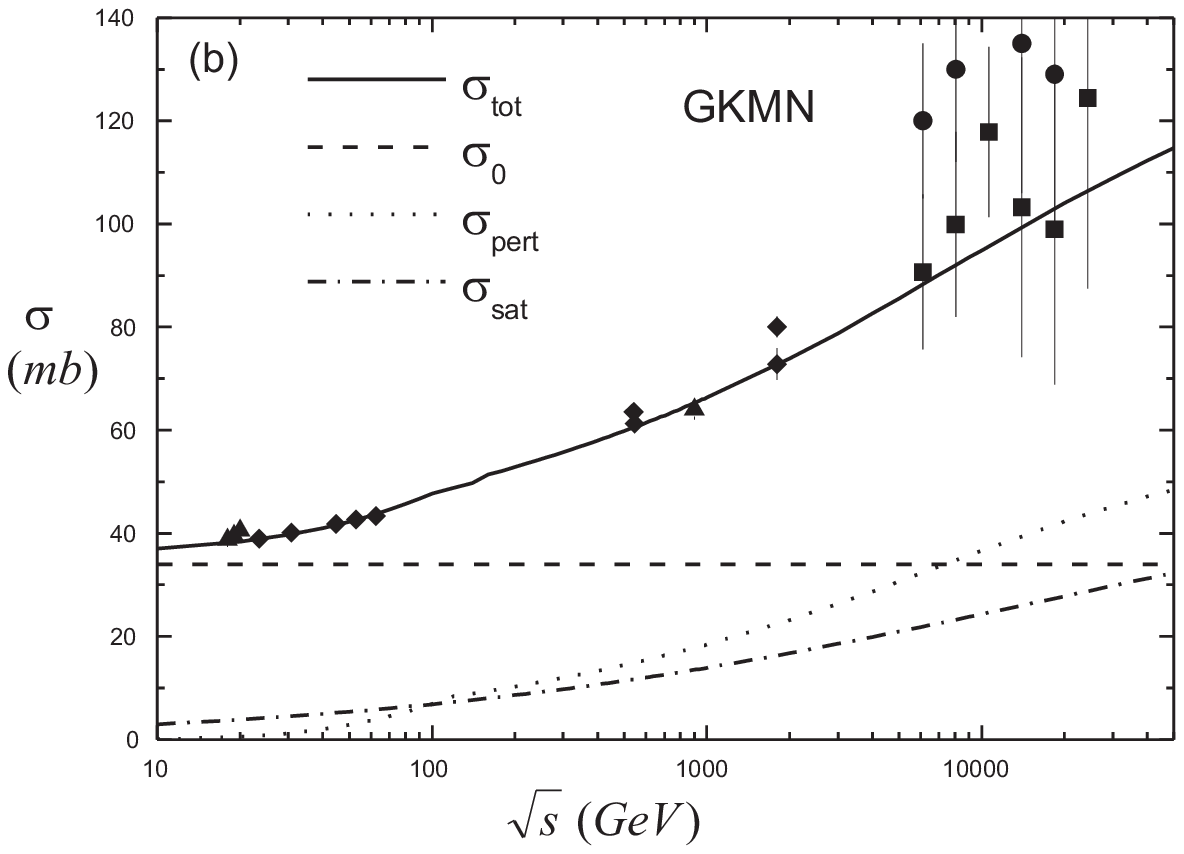}
\caption{Energy behavior of the total $pp/p\bar{p}$ cross section. The nonperturbative, 
perturbative and saturated components are  presented separately as well as  their sum, 
the total cross section. The results are for $\lambda = 0.25$. 
Data are from \cite{cern}, \cite{pp1},  \cite{pp1b}, \cite{pp2} and 
from \cite{pp3}.} 
\label{fig3}
\end{figure}

Following Ref. \cite{magnofl} we  consider a modification of the KKT model assuming that 
the saturation momentum scale is given by (\ref{qsgbw}) , $Y_0 = 4.6$, $c = 0.2$ and that 
the typical scale in the computation of $\xi (Y, r^2) $ is the photon virtuality. This 
modified model will be called KKTm.  
We also use the modified version \cite{rhic} of the DHJ model, called here GKMN, in which 
$Q_T = Q_0 = 1.0$ GeV, {\it i. e.} that the typical scale is energy independent.

In order to calculate $\sigma_{sat}$, it is also  necessary to specify the 
Bjorken-$x$ variable as in our previous calculations of $\sigma_{pQCD}$. We use the 
same prescription and the same value for $q_0$.

\section{Results}

In Fig. \ref{fig2} we show in arbitrary units the energy behavior of the
ratios $\sigma_{pQCD}/ \ln^2 s$ (solid lines) and
 $\sigma_{sat}/ \ln^2 s$ (dotted lines) for two choices of
$\lambda$. As it can be seen, all curves grow slower than $ln^2 s $. 
For smaller values of $\lambda$, such as 
$\lambda$ ($=0.1$) the fall of the  ratio shown in Fig. 2 would be postponed to 
very high energies,   $\sqrt{s} \simeq  10^6$ GeV. 
Although the energy at which the behavior 
of the cross section becomes ``sub-Froissart'' may depend strongly on $\lambda$, one 
conclusion seems very robust: {\it once $\lambda$ is finite, at some energy the growth of 
the cross section will become weaker than $ln^2 s$}.

After the study of the main properties of the perturbative and saturated components we  
can calculate the total cross section and compare with experimental data, 
obtained at CERN \cite{cern},  at Fermilab Tevatron \cite{pp1}, \cite{pp1b} 
and in cosmic ray experiments \cite{pp2,pp3}. The latter refer to proton-air cross sections 
and were translated to proton-proton cross sections in the phenomenological study of
Refs. \cite{bhs,niko}.

In Fig. \ref{fig3} we 
show the  sum $ \sigma_0 +\sigma_{sat} + \sigma_{pQCD}$ compared with the 
experimental data from Refs. \cite{cern,pp1,pp1b,pp2,pp3}.  $ \sigma_0$ was taken to 
be $34$ mb. In the figure the upper and lower panels were calculated with the KKTm 
and GKMN models, respectively.  The values of $q_0$ were $q_0=0.044$ GeV and  
$q_0=0.038$ GeV respectively.

Considering that there is 
only one free parameter ($q_0$) in our approach,  we obtain a good agreement with data.  
Moreover our predictions satisfy the Froissart bound. Probably a better agreement may be 
obtained if other quantities  are treated as free parameters, as for instance 
the effective exponent $\lambda$, and included  in a fitting procedure. 
\begin{figure}[t]
\includegraphics[scale=0.65]{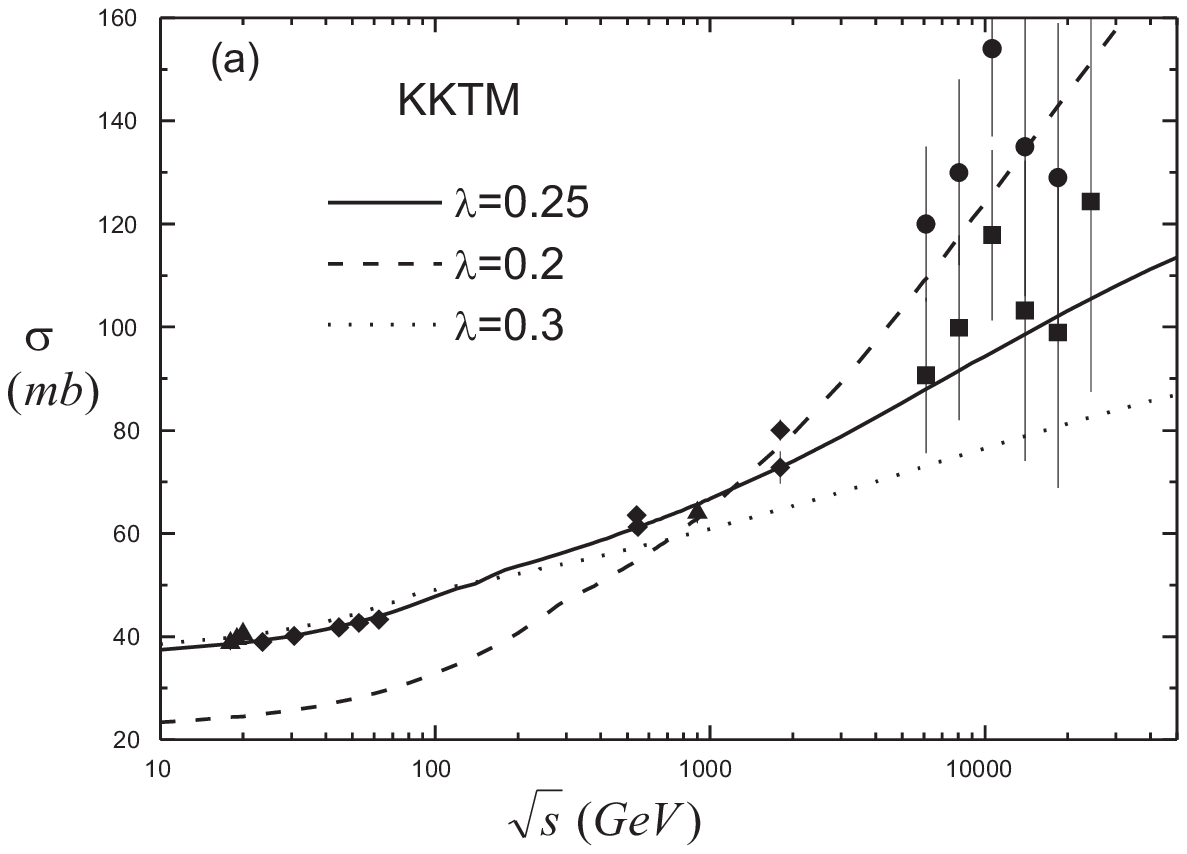}\\
\hspace{-0.1cm}
\includegraphics[scale=0.65]{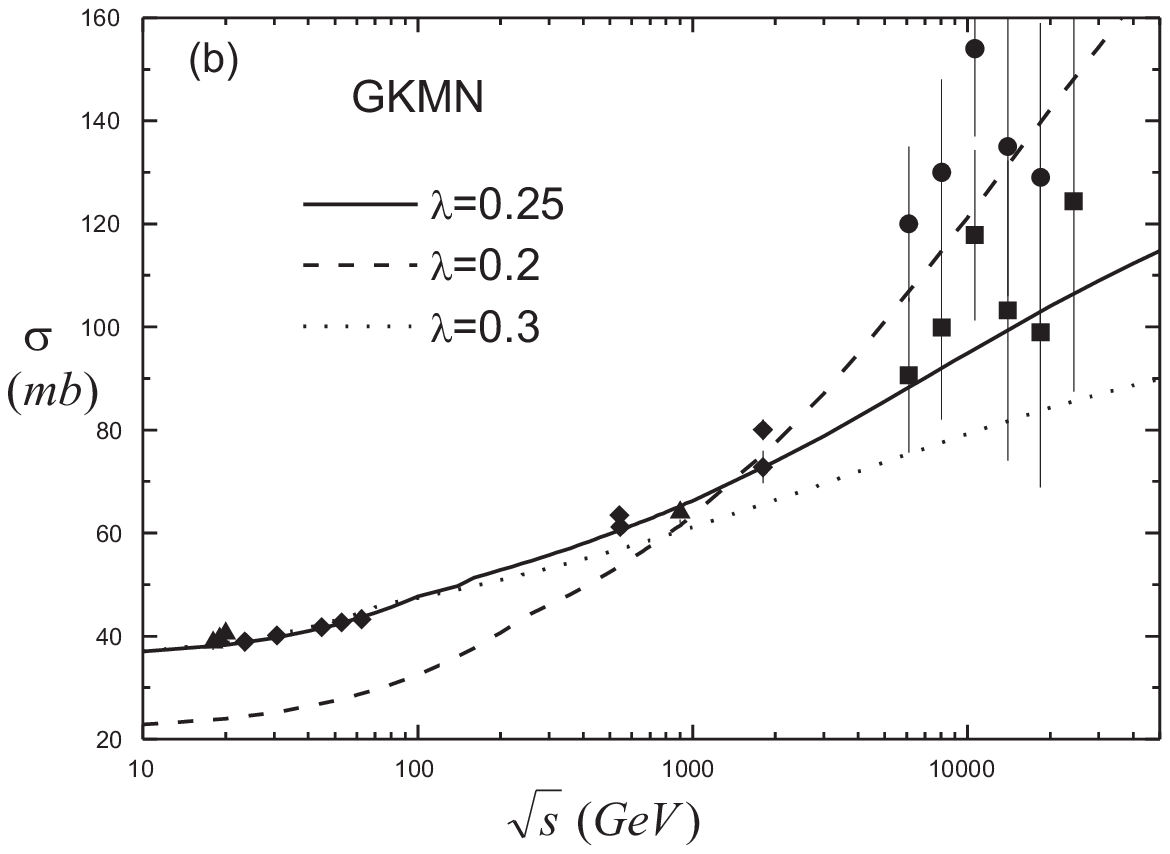}
\caption{Energy behavior of the total $pp/p\bar{p}$ cross section for  different values of 
the exponent $\lambda$. Data are the same as in Fig. 3. } 
\label{fig4}
\end{figure}
As already mentioned,  our results have a strong dependence on this quantity, as it can 
be seen in Fig. \ref{fig4}, where we present our results  for the total cross section 
for different values of $\lambda$. It is important to emphasize that there is only a small 
range of values of $\lambda$ which allow us to describe the experimental data. If, for 
instance,  
$\lambda = 0.4$  the  resulting  cross section  is very flat and clearly  below  
the data, while if $\lambda = 0.1$  the cross section grows very  
rapidly deviating strongly from the experimental data.  
The best choice for $\lambda$ is in the range $0.25 - 0.3$, which 
is exactly the range predicted in theoretical estimates using CGC physics and usually 
obtained by the saturation models for the DESY $ep$ HERA data.

\begin{figure}[t]
\includegraphics[scale=0.65]{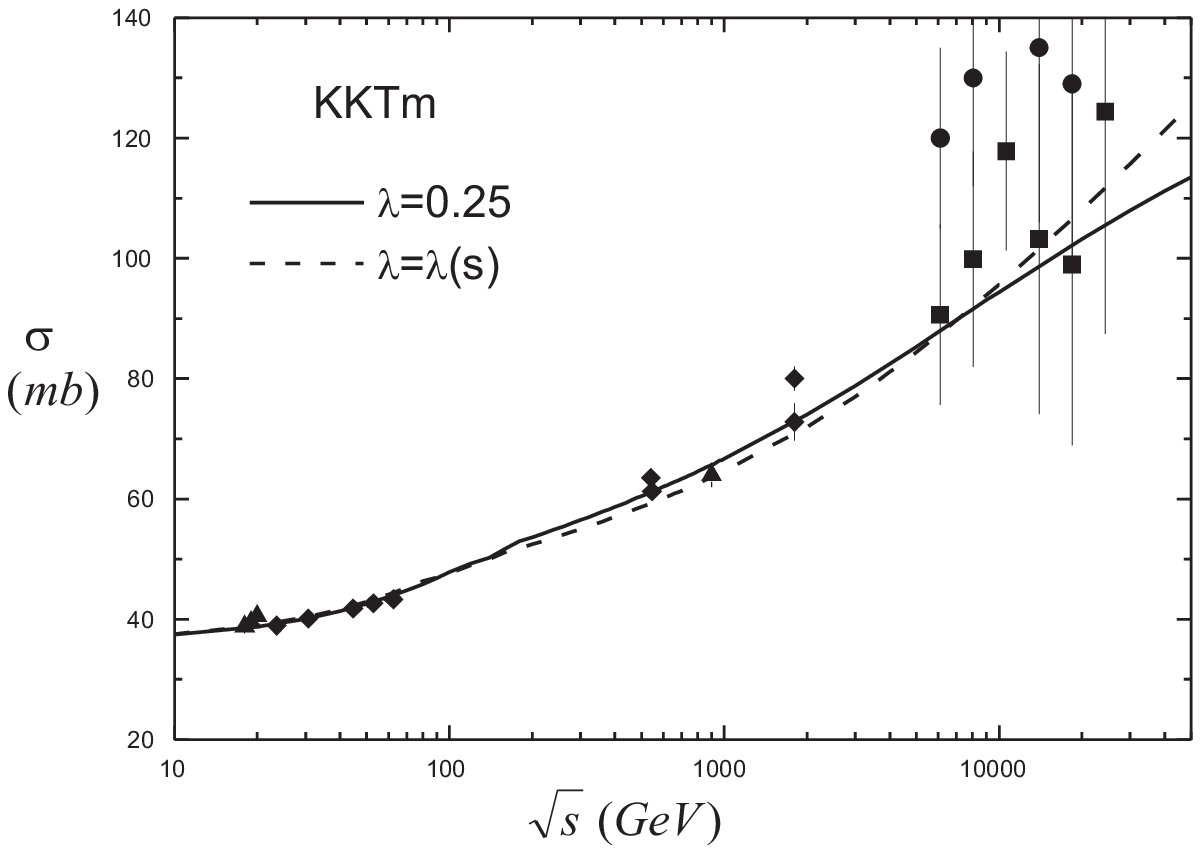}
\caption{Energy behavior of the total $pp/p\bar{p}$ cross section for  different 
values of the exponent $\lambda$. Data are the same as in Fig. 3. } 
\label{fig5}
\end{figure}

In the theory of the CGC the parameter $\lambda$ changes with the energy, being a function 
of the variable $Y= ln (1/x)$. Since our analysis is applied to a wide range of energies 
we have included the energy dependence of $\lambda$ as estimated in  \cite{trianta}, which 
can be parameterized as: 
\beq
\lambda = 0.3 - 0.003 (Y - 5)
\label{lambes}
\eeq
In Fig. \ref{fig5} we compare the cross sections obtained with a fixed value of $\lambda$ 
(= 0.25) and obtained with a ``running'' $\lambda$, according to (\ref{lambes}). As it 
can be seen, the difference between them is small.

\section{Conclusions}

In this paper we have proposed a simple model for the total 
$pp/p\bar{p}$ cross section, which is an improvement of the minijet model with the 
inclusion of a window in the $p_T$-spectrum associated to the saturation physics.
Our model implies a natural cutoff for the perturbative calculations which  modifies 
the energy behavior of this component, so that  it satisfies the Froissart bound. 
Moreover,  including the saturated component (calculated  with a dipole model), 
we obtain a satisfactory description of the experimental data. Our results  for the 
total $pp/p\bar{p}$ cross section also satisfy the Froissart bound. Finally, we find 
a very interesting consistency between our model and the saturation models used to 
describe the HERA data: similar values of $\lambda$ are needed to describe both set of 
experimental data. 

In other similar approaches, such as  \cite{bartels} the saturated 
cross section is used over the entire $p_T$ domain, or equivalently, for dipoles 
of all sizes. This procedure has two disadvantages: it requires the introduction of 
an (model dependent) impact parameter dependence of the dipole cross section and it 
does not make use of  the collinear factorization formula and  the parton densities, 
which, in the high $p_T$ region, are very well studied both theoretically and 
experimentally. In this sense our work is an improvement on \cite{halzen} and on 
\cite{bartels} as well.

\vspace{0.5cm} \underline{Acknowledgements}: This work was partially
 financed by the  Brazilian funding agencies FAPESP, FAPERGS and   CNPq.
\vspace{0.5cm}



\end{document}